\documentstyle[12pt,epsfig]{article}

\newskip\humongous \humongous=0pt plus 1000pt minus 1000pt

\newif\ifdtup

\def\gtap{\raisebox{-.4ex}{\rlap{$\sim$}} \raisebox{.4ex}{$>$}}

\begin{document}
\begin{titlepage}
\begin{center}
\today     \hfill    LBNL-39049 \\
~{} \hfill UCB-PTH-96/29  \\

\vskip .25in

{\Large \bf A Multilepton Signal For Supersymmetric Particles in Tevatron Data?}
\footnote{This work was supported in part by the Director, Office of 
Energy Research, Office of High Energy and Nuclear Physics, Division of 
High Energy Physics of the U.S. Department of Energy under Contract 
DE-AC03-76SF00098 and in part by the National Science Foundation under 
grants PHY-93-20551 and PHY-95-14797.}

\vskip 0.3in

R. Michael Barnett$^1$ and Lawrence J. Hall$^{1,2}$.

\vskip 0.1in

{{}$^1$ \em
     Lawrence Berkeley National Laboratory\\
     University of California, Berkeley, California 94720}

\vskip 0.1in

{{}$^2$ \em Department of Physics\\
     University of California, Berkeley, California 94720}

\end{center}

\vskip .3in

\begin{abstract}

The CDF and D0 collaborations have both reported unusual events in the
dilepton+jets sample with very high lepton and missing transverse
energies.  It is possible, but very unlikely, that these events
originate from top quark pair production; however, they have
characteristics that are better accounted for by decays of
supersymmetric quarks with mass in the region of 300 GeV:
$\widetilde q \rightarrow q \widetilde \chi$, $\widetilde\chi
\rightarrow \nu\widetilde\ell$, $\widetilde\ell\rightarrow \ell 
\widetilde \chi_1^0$.  Such a supersymmetric origin also leads to events with
large transverse missing energy and either 0, 1, 2 same-sign, or 3 isolated 
leptons.
\end{abstract}

\end{titlepage}

In Run I of the Tevatron Collider, the CDF and D0 collaborations announced the
discovery of the top quark in a variety of channels \cite{top}. Production of
$\bar{t} t$, with $t \rightarrow bW$ and both $W$s decaying leptonically leads
to $\ell^+ \ell^- jj \!\not\!\!E_T$ events where $\ell$ is an
energetic, isolated $e$ or $\mu$,
$\not\!\!E_T$ is missing transverse energy and $j$ represents a jet.
In 110 pb$^{-1}$, the CDF Collaboration observed 10 such opposite-sign dilepton
events, where 6 were expected from $\bar{t} t$ production with $m_t = 175$ GeV,
and 2 events were expected from non-top Standard Model backgrounds
\cite{conf,TARTARELLI,Kruse}. 

Two of these events, which we label $A$ and $B$, have characteristics which are 
quite unlike
those expected from $\bar{t} t$ production. While this could result from a 
statistical fluctuation of the top signal, we present Monte Carlo distributions
in Fig. 1 which illustrate how unlikely this is. The D0 collaboration has also
observed dilepton events, one of which
appears to have similar or even more dramatic characteristics \cite{Cochran}.

The dashed curves in Fig. 1 show Monte Carlo distributions for dilepton events
from $\bar{t} t$ production for $E_T^{l_1}, \not\!\!E_T, E_S$ and $\theta_T$.
$E_T^{l_1}$ and  $E_T^{l_2}$ are the transverse energies of the leading and
subleading leptons, $E_S =  E_T^{\ell_1} +  
E_T^{\ell_2} +  \not\!\!E_T$, and $\theta_T$ is the angle between the the two
isolated charged leptons in the plane transverse to the beam.
The $\theta_T$ distribution contains only events with $E_S > 250$ GeV. 

Event A has values for $E_T^{\ell_1}$, $\not\!\!E_T$ and $E_S$ which are 
on the tails of the distributions, see Fig. 1.
This event  contains a third isolated 
charged track, which is likely to be an
electron. A third isolated, hard charged lepton would make this event
inconsistent with a $\bar{t} t$ origin. In event B, 
the values for $E_T^{\ell_1}$, $\not\!\!E_T$ and $E_S$ are again high,
although the value for  $E_T^{\ell_1}$ is not as large as in event A. 
Fig.~1d shows that $\bar{t} t$ production with high $E_S$ leads
to large values of $\theta_T$. 
The measured values for $\theta_T$ are both
remarkably small, especially for event B. 
A kinematic argument 
shows that the values of  $E_T^{\ell_1}$, $E_T^{\ell_2}$ 
and $\not\!\!E_T$ of event B 
cannot arise from the decay of any pair of $W$'s whether or not 
the $W$'s originated in
$\bar{t} t $ production (neglecting neutrinos in the jets).

The central values of the variables $E_T^{\ell_1}$, $\not\!\!E_T$
and $E_S$ for the D0 event, labelled $C$, are far out on the
tails of the distributions. However, the large uncertainty in the 
measurement of the muon $E_T$ leads to large uncertainties in
these variables.
Although the event may be more dramatic than the CDF events,
this uncertainty means that a $\bar{t} t$ origin should not be excluded.
We believe that events A and B provide significant motivation 
for seeking an alternative origin for these dilepton events. 
Event C demonstrates that such exotic events may also be occurring in the D0
data.

The trademark of superpartner production at hadron colliders is 
well-known to 
be large $\not\!\!E_T$, signaling the escape of long-lived lightest
superpartners (LSPs) 
produced in the decays of supersymmetric particles. 
It is also
well-known that heavier squarks and gluinos tend to decay via a sequence of
cascades through charginos $(\widetilde{\chi}^+)$ and neutralinos  
$(\widetilde{\chi}^0)$, yielding events with isolated charged leptons $\ell$ 
as well as jets and $\not\!\!E_T$ \cite{THEORY,SS}. The isolated charged leptons
can arise from both $\widetilde{\chi}^+,\widetilde{\chi}^0$ decays, such as
$\widetilde{\chi}^+ \rightarrow \widetilde{\nu} \bar{e}$,
$\widetilde{e} \bar{\nu}$ and $\widetilde{\chi}^0 \rightarrow \widetilde{e} 
\bar{e}$, $\widetilde{\nu} \bar{\nu}$, and also from slepton decays, for 
example $\widetilde e\rightarrow e \widetilde\chi_1^0.$
The lepton carries high $E_T$, provided there is a large mass
difference between initial and final superpartners.
In this letter we point out that, in the minimal supersymmetric extension of 
the Standard Model, there are plausible ranges of superpartner masses in which
the cascade decays of squarks, $\widetilde{q}
\rightarrow  \widetilde{\chi} \rightarrow  \widetilde{\ell} \rightarrow \ell$,
could lead to a few  $\ell \ell (\ell) jj \!\not\!\!E_T$ events,
with extraordinarily high  $\not\!\!E_T$ and 
$E_T^\ell$, in the last Tevatron run. Such an interpretation of the events 
$A,B$ and $C$ appears reasonable, but only if the squarks have a mass of 
about 310 GeV -- 
beyond the reach previously thought possible for the Tevatron.

The minimal supersymmetric field content has two charginos and
four neutralinos, so there are many options for the nature of the relevant 
$\widetilde{\chi}$ states, which we call $\widetilde{\chi}'$,
and for the masses of the other $\widetilde{\chi}$ states. However,
in a 110 
pb$^{-1}$ run few heavy squarks and gluinos are produced,
so that the $\widetilde{\chi}'$ states must have a high branching
fraction to the desired leptonic mode. $\widetilde{\chi}$ states lighter than
$\widetilde{\chi}'$ might deplete the signal by allowing decays
$\widetilde{\chi}' \rightarrow \widetilde{\chi}H, \widetilde{\chi} W$, 
where $H$ and $W$ are Higgs and gauge bosons of any charge. 
This suggests that the other $\widetilde{\chi}$ states are heavier 
than the $\widetilde{\chi}'$, which we find to have a mass of about 260 GeV,
the exception being the lightest superpartner 
$\widetilde{\chi}_1^0$, which should be dominantly bino. 
The bino does not couple to $W$ and its
coupling to $H$ is proportional to the small hypercharge gauge coupling.

All charginos are at, or above, 260 GeV, so that the SU(2) gaugino mass
parameter $M_2 \;\gtap \; 260$ GeV. The left-handed slepton mass receives
radiative corrections from $M_2$, so that in all known models of supersymmetry
breaking these sleptons are heavier than 150 GeV. 
The charged $\widetilde{\chi}'$ are dominantly wino 
and therefore decay preferentially to left-handed sleptons.
The typical $E_T$ of the charged leptons from the cascade 
$\widetilde{\chi}' \rightarrow \widetilde{\ell}_L \ell$ is not as hard as the 
observed leading lepton in the collider events A, B and C. 
The most energetic leptons instead arise from slepton decay
$\widetilde{\ell}_L \rightarrow \widetilde{\chi}_1^0 \ell$,
so that the left-handed slepton mass must be large, in the
region of 220 GeV. 
The lepton spectrum is hardened if the mass of 
$\widetilde{\chi}_1^0$, the LSP, is small. This mass is given by the
hypercharge gaugino mass parameter, $M_1$, which is therefore several times less
than $M_2$.
The region of parameters of interest to us does not
allow the relation $M_2 \approx 2 M_1$, which occurs in simple
schemes of grand unification with large messenger scales for supersymmetry
breaking.

What are the gaugino/higgsino components of the $\widetilde{\chi}'$ states?
If they were dominantly higgsino, they would 
be produced only in the decay of
$\widetilde{t}_R$, since other flavors of squarks would decay predominantly to 
the LSP bino, $q \widetilde{\chi}^0_1$. 
To match the observed event rate would require a lower
mass for $\widetilde{t}_R$, and therefore of $\widetilde{\chi}'$ and 
$\widetilde{\ell}$, softening the $E_T^\ell$ and $\not\!\!E_T$ distributions. 
Furthermore, since the $\widetilde{\chi}'$ states would decay to 
$\widetilde{\ell}\ell$ only through their small wino
components, this leptonic branching ratio would be depleted by the decay 
$\widetilde{\chi}' \rightarrow H \widetilde{\chi}^0_1$, 
at least for the neutral $\widetilde{\chi}'$ states 
where the $H^0$ is guaranteed to be light. We conclude
that the $\widetilde{\chi}'$ states most likely 
have substantial gaugino components. In
this letter we study the simplified case where there are three 
$\widetilde{\chi}'$ states ($\widetilde\chi$ states relevant to
$\widetilde q$ decays), and they are dominantly the $SU(2)_L$ gauginos: 
$\widetilde{\chi}_1^{\pm} \approx \widetilde{w}^{\pm}$ and  
$\widetilde{\chi}_2^0 \approx \widetilde{w}_3$. This is achieved
by making the Higgsino mass parameter $|\mu| > 400$ GeV (for $M_2 = 260$ GeV). 
The heavier 
states $\widetilde{\chi}_2^{\pm}$ and $\widetilde{\chi}_{3,4}^0$ have masses
near
$|\mu|$ and play no role in our analysis. Similarly, the small 
Higgsino components of the $\widetilde{\chi}'$ states are unimportant. We stress
that while large $|\mu|$ was chosen for simplicity, the other aspects of the
superpartner spectrum were dictated by the requirement of hard
$E_T^\ell$ and $\not\!\!E_T$ distributions.

The requirement of hadron collider events with sufficiently hard 
$E_T^\ell$ and $\not\!\!E_T$ distributions has led us to a remarkably simple
and plausible parameter region of the minimal extension of the supersymmetric
standard model. There are five flavors of left-handed squarks with masses in 
the region of 310 GeV. These decay to $SU(2)_L$ gauginos,
$\widetilde{\chi}_1^{\pm}$ and $\widetilde{\chi}^0_2$, 
of mass near 260 GeV, which in turn decay to left-handed 
sleptons with mass near 220 GeV. The hardest charged leptons are
produced in the 
final cascade of the sleptons to the LSP $\widetilde{\chi}^0_1$, which is
dominantly bino. 
With a squark mass as high as 310 GeV, it is remarkable that present collider
data may contain a few events from this cascade: 
$\widetilde{q}_L \rightarrow \widetilde{\chi}_1^+,\widetilde{\chi}_2^0 
\rightarrow \widetilde{\ell}_L \rightarrow \ell \widetilde{\chi}_1^0$.

The right-handed slepton masses are expected to be substantially 
smaller than the 
left-handed slepton masses because they do not receive the large radiative 
correction from $M_2$. This is fortunate since otherwise the LSP,
$\widetilde{\chi}_1^0$,
would overclose the universe. Bino dark matter gives $\Omega h^2 = 0.5$,
for $M_1 = 50$ GeV with three degenerate right-handed sleptons of mass
$m_{\widetilde{\ell}_R} = 130$ GeV.

The two top squarks, $\widetilde{t}_{1,2}$, may be lighter or heavier than the
other squarks. If heavier, they are rarely produced and are irrelevant. If
lighter they can add to the hard lepton signature, by decaying to
$\widetilde{\chi}' q$, or to the dijet $+ \not\!\!E_T$ signature by 
decaying to $t \widetilde{\chi}_1^0$. 
We have not included such possible contributions in our analysis.

In our scheme, dilepton events, such as events $A$ and $B$, arise
from the decay of
a $\widetilde{q}_L^{(\dagger)} \widetilde{q}_L$ pair.
Since $\widetilde{g}$ decays to $ q^\dagger \widetilde{q}$ or to
$q \widetilde{q}^\dagger$, events from $\widetilde{g} \widetilde{q}$ and
 $\widetilde{g} \widetilde{q}^\dagger$ production (which can occur at a larger
 rate than  $\widetilde{q}^\dagger \widetilde{q}$ events) look similar to  
$\widetilde{q}^\dagger \widetilde{q}$ events.
Because $\widetilde{q}_L$ has a small hypercharge, the direct decay 
$\widetilde{q}_L \rightarrow q \widetilde{\chi}^0_1$ has a small branching
ratio compared to the cascade mode $\widetilde{q}_L \rightarrow 
\widetilde{\chi}^0_2, \widetilde{\chi}^+_1 \rightarrow \widetilde{\ell}
\rightarrow \ell \widetilde{\chi}^0_1$, 
even for our choices of masses $(m_{\widetilde{q}_L} =
310 \mbox{ GeV}, m_{\widetilde{\chi}_1^+}= m_{\widetilde{\chi}_2^0}= 260$ GeV)
where the cascade mode is phase space suppressed.  
The predominant decay of each squark is therefore
$\widetilde{q}_L \rightarrow q (\widetilde{\chi}_1^\pm, \widetilde{\chi}^0_2)$,
followed by $\widetilde{\chi}^-_1 \rightarrow \ell' \widetilde{\nu}^\dagger,
\bar{\nu} \widetilde{\ell}$ and $\widetilde{\chi}^0_2 \rightarrow \ell' 
\widetilde{\ell}^\dagger, \nu \widetilde{\nu}^\dagger$, (together with the
Hermitian conjugate decays), and finally $\widetilde{\ell} \rightarrow \ell
\widetilde{\chi}_1^0$ and  $\widetilde{\nu} \rightarrow \nu
\widetilde{\chi}_1^0$. 
The $E_T^\ell$ distribution for the $\ell'$, which appears in 
$\widetilde{\chi}'$ decay, is softer than for the $\ell$, which appears in 
$\widetilde{\ell}$ decay (for the masses we derive). 
These events therefore have a wide range of characteristics, with the number of
isolated charged leptons, $N_L$, varying from 0 to 4, as shown in
the Table.

\newpage
\vskip 20pt
\centerline {\bf Table}
\begin{center}
\begin{tabular}{|c|c|c|c|}
\hline
$N_L$& $\widetilde{\chi}'$ charges&Charged leptons&Relative rate\\
\hline \hline
0&(0,0)&&1\\
\hline
1&$(\pm,0)$&$\ell'$&4\\
1&$(\pm,0)$&$\ell $&4$\rho$\\
\hline
2&(+,--)\ ($\pm$,$\pm$)&$\ell' {\ell}'$&4\\
2&(0,0)\ (+,--)\ ($\pm$,$\pm$)&$\ell' {\ell}$&10$\rho$\\
2&(+,--)\ ($\pm$,$\pm$)&$\ell {\ell}$&4$\rho^2$\\
\hline
%3&$(\pm,0)$&$\ell' \ell' \bar{\ell}  + \bar{\ell}' \bar{\ell}' \ell$&2$\rho$\\
3&$(\pm,0)$&$\ell' \ell'  {\ell}$&4$\rho$\\
%3&$(\pm,0)$&$\ell' \ell  \bar{\ell}  + \bar{\ell}' \bar{\ell} \ell$&2$\rho^2$\\
3&$(\pm,0)$&$\ell'  \ell  {\ell}$&4$\rho^2$\\
\hline
%4&(0,0)&$(\ell' \bar{\ell})^2 +(\bar{\ell}' \ell)^2$&${1 \over 2}\rho^2$\\
4&(0,0)&$\ell' {\ell}'  \ell {\ell}$&$\rho^2$\\
\hline
\end{tabular}
\end{center}     

The second column labels the charges of the two $\widetilde{\chi}'$ states
produced from the two squark decays. The third column lists the possible 
combinations of the isolated leptons.  Each
$\ell$ stands for $e$, $\bar{e}$, $\mu$, or $\bar{\mu}$, which occur with
equal probability. The flavors of different leptons are uncorrelated,  except 
for those arising from $\tilde{\chi}^0_2 \rightarrow \ell'
\widetilde{\ell}^\dagger \rightarrow \ell' \bar{\ell}$, 
when the flavors of the lepton and anti-lepton are identical.
For simplicity we ignore the additional events with isolated charged leptons
which can occur when one or both $\widetilde{\chi}'$ states decay to the $\tau$
flavor. 

The relative rates for the event categories are shown in the last column of the
Table. The phase space factor $\rho = (1 - m_{\widetilde{\ell}}^2/M_2^2)^2
(1 - m_{\widetilde{\nu}}^2/M_2^2)^{-2}$ is less than unity since the $D$ terms
give $m_{\widetilde{\nu}} < m_{\widetilde{\ell}}$. For $\rho = 1$, the Table
shows that $3\over 4$ 
of these $\widetilde{q}_L^{(\dagger)} \widetilde{q}_L$ events have
$N_L \geq 2$. Since the $\ell'$ are typically soft, some care is necessary in
interpreting these events. For example, the ``mixed" dilepton events, $\ell'
\bar{\ell}$, typically have one very hard and one softer lepton, as observed for
event A.

If some of the dilepton events at the Tevatron Collider have a
supersymmetric origin, there is certainly an overlap region between
these events and those from top quark decays.  We have identified three
events that seem very unlikely to originate from top decays.
However, it is possible that there are other events whose
characteristics are somewhat less dramatic but also come from squark
decays. 

Since each event would involve two separate squark decays, we can
learn a little about the distributions by comparing the two lepton
$E_T$.  Event A may be instructive, since it contains both
a 182 GeV lepton and a 27 GeV lepton.  It is possible that one is at
the high end of the distribution and the other at the low end.

With only three identified candidates for squark decays, it is not
possible to quantitatively describe any distributions.  Furthermore,
the muon in event C (with $E_T\approx 200$ GeV) is not
well-measured.
Nonetheless it is possible to identify approximate masses of six
supersymmetric particles within the assumed scenario.  Given the
small number of events, we will not give error bars on these masses,
but estimate them as about $\pm (10-30)$ GeV.

The left-handed squark mass is determined to be about 310 GeV by the observed 
event rate, the production rate for $\widetilde{q}_L^\dagger \widetilde{q}_L$
and  $\widetilde{g} \widetilde{q}_L^{(\dagger)}$, 
and the successive branching ratios, as discussed later.  
The squark decay $\widetilde q_L \rightarrow q \widetilde \chi'$ yields
jets with $E_T\sim $ 20-80 GeV though mostly at the low end.  The
implied $\widetilde q_L - \widetilde \chi'$ mass splitting gives
$ m(\widetilde\chi_1^+) \sim  m(\widetilde\chi_2^0) \sim 260$ GeV.

In order to achieve the highest observed lepton $E_T$, it is
essential that the sleptons have a very high mass and that
$\widetilde \chi_1^0$ not be too heavy.  To be consistent with 
cosmology, and to maintain a
large mass splitting with the sleptons, we take
$m(\widetilde\chi_1^0) \sim 50$ GeV.  To obtain the hard lepton
spectrum and not produce soft dilepton events (which are not seen in
excess of the top quark events), the highest possible slepton mass
is needed $m(\widetilde\ell) \sim m(\widetilde\nu) \sim$ 210-230
GeV.

Having estimated the masses of $\widetilde q$, $\widetilde\ell$,
$\widetilde\nu$, $\widetilde \chi_2^0$, $\widetilde\chi_1^+$, and
$\widetilde\chi_1^0$, we plot several distributions in Fig.~1. 
Since these events all have $E_T^{\ell_1}+ E_T^{\ell_2}
+ \not\!\!E_T$ well above 250 GeV, we
use this cut in examining the transverse angle between the leptons
(in Fig.~1d).
Not shown but also interesting is the distribution of the slower lepton
in the mixed mode which can yield softer leptons in one part of an
event in some cases.

\begin{figure}
%\begin{center}
\epsfig{file=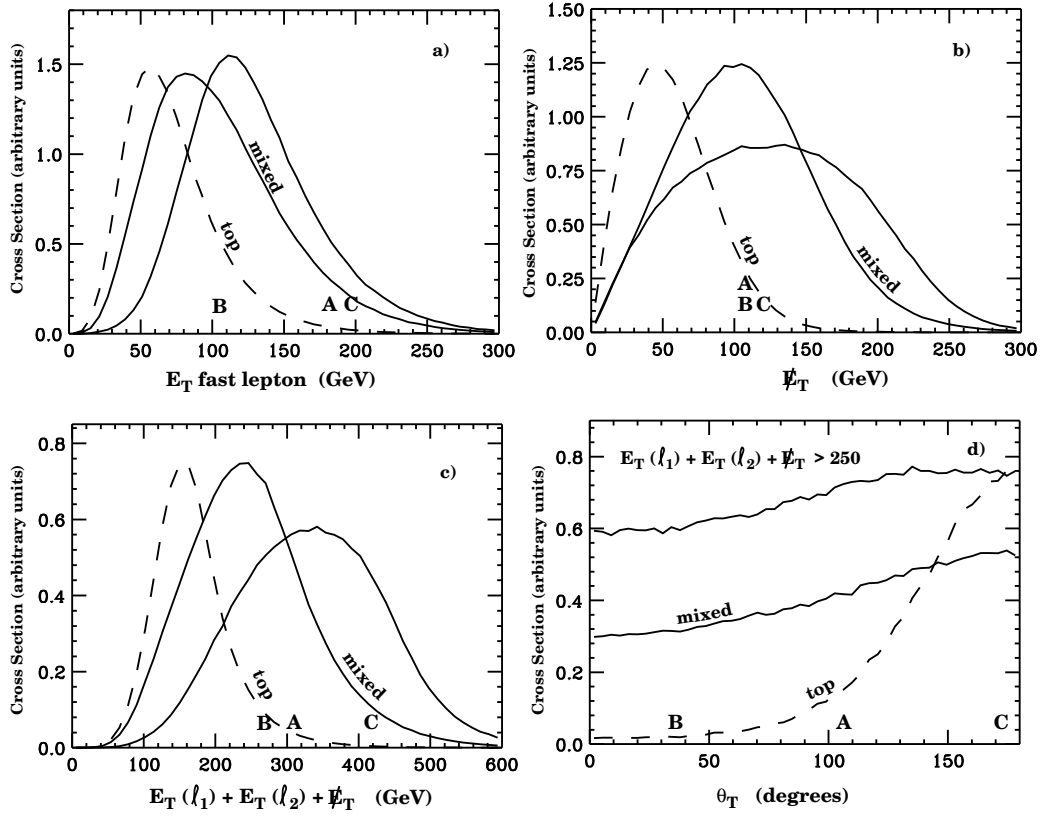,width=\textwidth}
%\end{center}
\caption{Expected distributions for (a) $E_T^{\ell_1}$ , (b) $\not\!\!E_T$ , 
(c) $E_T^{\ell_1}+E_T^{\ell_2}+ \not\!\!E_T$ , and (d) 
$\theta_T$ (between the two leptons).  The dashed curves are for
$t\bar t$  production with $m_t=175$ GeV.  The curve labelled
``mixed'' has one lepton from $\widetilde\chi^\prime \rightarrow
\ell \widetilde\nu$ or $\ell\widetilde\ell$ and one from $\widetilde
\ell \rightarrow \ell \widetilde\chi_1^0$.  The other solid curve
has both leptons from $\widetilde \ell \rightarrow \ell 
\widetilde\chi_1^0$ decays.  The three events mentioned in the text
are labelled A, B, and C.  Note that event A also has an isolated
track that may be a third energetic lepton;  this would increase the
sum of $E_T$ for this event in (c).  The $E_T$ of the muon in event
C is very poorly measured; this uncertainty impacts on (a), (b), and
(c). }
\label{fig:1}
\end{figure}

In our scheme, events $A$ and $B$ arise from cascade decays of 
$\widetilde{q}_L^{(\dagger)} \widetilde{q}_L$. In a run of 110 pb$^{-1}$ the
expected number of events with $N_L \geq 2$ is ($\sigma_T / 0.05$ pb)
($\epsilon /0.25$), where $\epsilon$ is the detection efficiency,
which we estimate to be 0.25, and $\sigma_T$
is the total 
$\widetilde{q}_L^\dagger \widetilde{q}_L+
 \widetilde{q}_L \widetilde{q}_L+
 \widetilde{q}_L^\dagger \widetilde{q}_L^\dagger$ production cross
section. There are two contributions to $\sigma_T$ which may be important:
direct $\widetilde{q}_L^\dagger \widetilde{q}_L$ production, and
$\widetilde{q}_L \widetilde{g}$, $\widetilde{q}_L^\dagger \widetilde{g}$ 
production followed
by $\widetilde{g} \rightarrow \widetilde{q}_L^\dagger q, \widetilde{q}_L
q^\dagger$.
The relative importance of these two contributions depends on
$m_{\widetilde{g}}$ and $m_{\widetilde{q}_R}$, which we have not
determined\footnote{We do not assume the universality relation  
$m_{\widetilde{q}_R} =  m_{\widetilde{q}_L}$, nor do we exclude it.}. For
example, with  $m_{\widetilde{g}} = 330$ GeV and  
$m_{\widetilde{q}_R} = m_{\widetilde{q}_L} = 310$ 
GeV, the direct production contributes 0.03 pb to $\sigma_T$, while
squark-gluino production contributes $0.07$ pb to $\sigma_T$.
For these parameters, a further production rate, $\sigma B$, for
dilepton events of 0.04 pb arises from 
$\widetilde{q}_L^{(\dagger)} \widetilde{q}_R^{(\dagger)}$
production, giving a total expectation of about 3 events with
$N_L\geq 2$.  If $\widetilde\tau_L$ is degenerate with 
$\widetilde e_L$ and $\widetilde\mu_L$, this would be depleted by a
factor of about 2.

For many years it has been proposed that a primary signature for
supersymmetry would be events with leptons, jets, and missing
energy.  It was always clear that top quark decays would be a major
background.  We cannot demonstrate that the events we have studied
are due to squark decays, but they appear not to come from
top decays.

Perhaps the most notable result of our analysis is that with only
three candidate events from a hadron collider, we are able to
roughly estimate the masses of six supersymmetric particles (and the
gaugino/Higgsino content of the $\widetilde\chi'$ states at 260 GeV).  
Clearly more data are needed to refine these estimates and to establish the
particular scenario we have described.

The reported rate for dilepton events is about 30\% higher than that
expected from top quarks, but this excess is not statistically
significant. 
If our scenario is correct, we also anticipate the observation of other
types of events (though some may have significant backgrounds).  
As shown in the Table, we
expect events with large missing $E_T$ and 0, 1, 2, 3, and (very
rarely) 4 leptons. These 1-lepton events may have only two jets  
and hence would not be in the top quark sample. The trilepton events are
expected to contain jets, unlike those which would result from the
production of much lighter chargino-neutralino pairs.

Additional signatures may
also be present, depending on the values of $m_{\widetilde{q}_R}$ and 
$m_{\widetilde{g}}$. The production of $\widetilde{g}\widetilde{q}$ contributes
equally to same-sign \cite{SS} and opposite-sign dileptons 
($\widetilde{q}^{(\dagger)} \widetilde{q}$ production can also lead to same-sign
events). When right-handed squarks are
produced, they decay directly to the LSP: $\widetilde{q}_R \rightarrow q_R
\widetilde{\chi}^0_1$, so that several new signals are possible. For example,
with $m_{\widetilde{g}} = 330$ GeV and $m_{\widetilde{q}_R} = 
m_{\widetilde{q}_L} = 310$ GeV, we find a production rate,
$\sigma B$, for 
($jj \not\!\!E_T,jj\ell \not\!\!E_T$) 
events of (0.15, 0.19) pb.
The standard model backgrounds for these $N_L = 0,1$
events are larger than for the case of $N_L = 2$. However, the signal events
are prominent: the $jj \not\!\!E_T$ events have $E_T^j \sim50-230$ GeV and
$\not\!\!E_T \sim 50-280$ GeV.
For values of $|\mu|$ below 400 GeV, the decays of charginos to $W^+
\widetilde{\chi}_1^0$, $H^+ \widetilde{\chi}_1^0$
and neutralinos to $Z \widetilde{\chi}_1^0$, $H^0 
\widetilde{\chi}_1^0$ may become important.

Remarkably we find that the Tevatron Collider experiments can be
sensitive to very high squark masses, in excess of 300 GeV.  
In our scenario this happens because the relevant charginos
and neutralinos have masses between the squarks and the sleptons,
leading to high leptonic branching ratios and to hard $E_T^\ell$ and
$\not\!\!E_T$ distributions. The reach in squark mass exceeds that of several
previous analyses, because the signal can be distinguished from the
$\bar{t} t$ background. Without such a kinematic distinction, the signal can
only be seen with large statistics, leading to a more limited reach in the
squark mass.
The superpartner masses of our scheme
are so high that no supersymmetric particle would be found at LEP2,
and a 500 GeV NLC would not find all of these particles. If this
turns out to be the first evidence for supersymmetry, the
confirmation will come in the next Tevatron run which may obtain
10-20 times as many events. 
It may also be possible to identify a few events
with large $\not\!\!E_T$ and 0, 1, 2 same-sign, or three isolated leptons in the
present data.

\begin{center}               
{\bf Acknowledgments} 
\end{center}

We thank members of the CDF and D0 collaborations for useful conversations.

%{\em This work was supported by the Director, Office of Energy Research,
%Office of High Energy and Nuclear Physics, Division of High Energy 
%Physics of the U.S. Department of Energy under Contract DE-AC03-76SF00098.}
This work was supported in part by the Director, Office of 
Energy Research, Office of High Energy and Nuclear Physics, Division of 
High Energy Physics of the U.S. Department of Energy under Contract 
DE-AC03-76SF00098 and in part by the National Science Foundation under 
grants PHY-93-20551 and PHY-95-14797.

% '99' is at least as wide as the widest bibliography label, 
% could use '9' if there are less than 10 references. 


\begin{thebibliography}{99}
\bibitem{conf} These events have been shown at recent conferences, such as
SUSY96, and are included in the reported top quark dilepton sample.
\end{thebibliography}
\end{document}